\documentclass[12pt ]{article}
\usepackage[dvips]{epsfig}
\newcommand{\ee}{$e^+e^-\: $}

\newcommand{\qq}{$q\,\bar{q}\:\;$}

\newcommand{\nn}{$\nu\bar{\nu}\: $}

\newcommand{\gaga}{$\gamma \gamma\:\;$}
\newcommand{\hgg}{$H \rightarrow \gamma \gamma  \:$}
\newcommand{\noi}{\noindent}

\begin{document}

\pagestyle{empty}


\noi \hspace*{14cm}LC-PHSM-2000-053\\
\noi \hspace*{14.5cm}November 2000\\

\vspace{2.0cm}
\begin{center}
\LARGE{\bf\boldmath Measuring the Higgs Branching Fraction into two
  Photons at Future Linear \ee Colliders \unboldmath}
\end{center}

\vspace{0.8cm}
\large
\begin{center}
E. Boos$^1$, J.-C. Brient$^2$, D.W. Reid$^3$, H.J. Schreiber$^4$,
 R.Shanidze$^5$
\end{center}

\vspace{0.3cm}
\bigskip \bigskip  
\begin{center}
\small
$^1$ Institute of Nuclear Physics, Moscow State University, \\
119 899 Moscow, Russia \\ [2mm]

$^2$ Laboratoire de Physique Nucl\'{e}aire et des Hautes Energies,
Ecole Polytechnique, IN$^2$P$^3$-CNRS,
 F-91128 Palaiseau Cedex, France \\ [2mm]

$^3$ NIKHEF, Postbus 41882, 1009 DB Amsterdam, The Netherlands \\ [2mm]

$^4$ DESY Zeuthen, 15735 Zeuthen, FRG \\ [2mm]

$^5$ High Energy Physics Institute, Tbilisi State University,
380086 Tbilisi, Georgia; \\
now at Physikalisches Institut, Universit\"at Erlangen-N\"urnberg,
91058 Erlangen, FRG
\normalsize
\end{center}

\vspace{1.5cm}
\pagestyle{plain}
\pagenumbering{arabic}

%
\section*{Abstract}

We examine the prospects for measuring the \gaga branching fraction of a
Standard Model-like Higgs boson with a mass of 120 GeV
at the future TESLA linear
\ee collider, assuming an integrated luminosity of
1 ab$^{-1}$ and center-of-mass energies of 350 GeV and
500 GeV. The Higgs boson is produced in association with a fermion pair
via the Higgsstrahlung process \ee $\rightarrow ZH$,
with $Z \rightarrow$ \qq or \nn, or the 
WW fusion reaction $e^+e^- \rightarrow \nu_e \bar{\nu_e} H$.
A relative uncertainty on BF(\hgg) of~16\% 
can be achieved in unpolarized \ee
collisions at $\sqrt{s}$=~500~GeV, while for
$\sqrt{s}$=~350~GeV the expected precision is slightly poorer.
With appropriate initial
state polarizations $\Delta$BF(\hgg)/BF(\hgg) can be improved
to 10$\%$. If this measurement is combined with the
expected error for the total
Higgs width, a  precision of 10\% on the \gaga Higgs boson partial width
appears feasible.

\newpage

\section {Introduction}

 Following the discovery of the Higgs boson, one of the main tasks
 of a future linear
 \ee collider will be precise model-independent measurements
 of its fundamental
 couplings to fermions and bosons and its total
 width \cite{LC1}.
 The branching fraction of the Higgs boson into two
 photons, BF(\hgg), is of special interest since
 deviations of BF(\hgg) 
 (or the diphoton Higgs partial width $\Gamma$(\hgg))
 from the Standard Model (SM) value provide sensitivity to
 new physics. In particular,
 by virtue of the fact that the \hgg coupling can have contributions from loops
 containing new charged particles, significant differences from the SM
 value are possible.
 Thus, a measurement of BF(\hgg) 
 at the next linear collider will be an important contribution
 to understanding the nature of the Higgs boson and may
 possibly provide hints for new
 physics, regardless of the size of the deviation from the SM prediction. 
 The precision of the $H \rightarrow \gamma \gamma$  branching fraction
 measurements within reach in \ee collisions
 is the subject of this paper.         

 The ultimate goal of this measurement is to derive the
 diphoton Higgs boson partial width.  However this
 needs knowledge
 of the total Higgs width.
 The total width of a light SM Higgs boson
 is too small to be observed  directly~\cite{GH}.
 The procedure for obtaining this quantity
 requires measurements of
 the product ${\Gamma(H\rightarrow \gamma \gamma)
\cdot BF(H\rightarrow b \bar{b})}$
 in the $\gamma\gamma$ collider option of a linear
 collider and $BF(H \rightarrow b \bar{b})$, a quantity easily
 accessible with high precision in \ee collisions \cite{HJS1}.               
 As both these measurements can be achieved with an accuracy
 of few percent,
 the error of the total Higgs width will be dominated by the
 error of  $BF(H \rightarrow \gamma \gamma$).

However, it has recently been demonstrated that a measurement of the
branching fraction BF$(H \rightarrow W W^{*})$
\cite{Bo} at a high-luminosity linear \ee collider,
combined with a precise value for the rate of the WW
fusion process $e^+e^- \rightarrow \nu_e \bar{\nu_e} H$
(or for the Higgsstrahlung production rate $\sigma(HZ)$
 and assuming W,Z-universality),
would permit an accurate measurement of the 
total width of the Higgs boson.

Precise electroweak data indicate the existence of a light Higgs
boson \cite{La} with a mass below about 200 GeV, with a preference
for $M_H$ close to 120 GeV.
High-luminosity linear \ee colliders in the energy range
300 to 500 GeV \cite{CDR} are ideal
machines for performing precise measurements
of the properties of such a particle.
In this paper we investigate the prospects of measuring the branching
fraction BF(\hgg) from events of the reactions
\begin{equation}
       e^+e^- \rightarrow q \bar{q}\gamma\gamma
\end{equation}
  and
\begin{equation}
       e^+e^- \rightarrow \nu \bar{\nu}\gamma\gamma
\end{equation}
assuming a Higgs boson mass $M_H$ = 120 GeV,
at $\sqrt{s}$ = 350 and 500 GeV
and an integrated luminosity of 1 ab$^{-1}$ at each energy.

The statistical precision for
BF(\hgg) is mainly determined by
$\sqrt{S+B} /S$, where S and B are respectively the
number of signal and background events 
within a small interval of the
two-photon invariant mass $\Delta M_{\gamma \gamma}$,
centered around $M_H$. Hence, evaluation
of all relevant signal and background processes and
optimization of selection procedures are mandatory,
taking into account acceptances and resolutions of a linear collider
detector.

Our analysis is superior in some respects to the study of
ref.\cite{Reid}. It includes the complete irreducible background
in reactions (1) and (2) and demonstrates for the first time the gain
in the precision of BF(\hgg) when beam polarization is accounted for
in signal and background events.

 The paper is organized as follows. In Section~2 we discuss
 simulation of the Higgs signal and background events and
 their detector response. In Sections~3 and~4 we present
 our results for unpolarized BF(\hgg) measurements at
 $\sqrt{s}$~=~350 and 500~GeV, respectively. 
 In Section~5 we discuss
 improvements to the \hgg branching fraction measurement
 with beam polarization.
 Section~6 summarizes the conclusions.

\vspace{1.0cm}
\section {Signal and Background Reactions}

 In $e^+e^-$ collisions the Higgs boson is produced by two different
 processes, the Higgsstrahlung process
\begin{equation}
      e^+e^- \rightarrow Z H 
\end{equation}
and the weak boson (WW and ZZ) fusion reactions \\
\begin{equation}
      e^+e^- \rightarrow \nu_e \bar{\nu_e} H   
\end{equation}
\begin{equation}
      e^+e^- \rightarrow e^+ e^- H
\end{equation}
   
 The ZZ fusion process (5) is strongly suppressed with
 respect to reaction (4) (by about a factor of 10
 relatively independent of $\sqrt{s}$). Therefore,
 only the Higgsstrahlung and WW fusion reactions
(3) and (4) are considered in this study.
These processes are part of the 2-to-4 body reactions
(1) and (2) if only the most important
Z $\rightarrow q\bar{q}$ (of $\sim$70$\%$) and
$\nu \bar{\nu}$ (of $\sim$20$\%$) decays and the \hgg decay are
accounted for. Thus, to be most general in our analysis, events
of reactions (1) and (2) were generated
 by means of the program package CompHEP \cite{CompHEP}, including
 initial state bremsstrahlung and beamstrahlung for the TESLA linear
collider option \cite{bms}. In this way, Higgs boson production
and the complete irreducible background as well as possible
interferences are taken into account.
 The branching fraction for \hgg was imported from the program
 package HDECAY \cite{HDECAY}. It depends on the Higgs mass and
 is largest near $M_H$ = 120 GeV. In this study we used
 BF(\hgg)= $2.2 \times 10^{-3}$.

The Higgsstrahlung reaction (3) is characterized by two hadronic jets
originating from the Z, together with two energetic photons
with an invariant mass equal to $M_H$. The background
expected in reaction (1)
comes from the 100 lowest-order diagrams
with $q=d,u,s,c,b$, which are shown, for the d-quark as
an example, in Fig.~1. The most serious background arises from
the double bremsstrahlung process \ee $\rightarrow
Z \gamma \gamma \rightarrow q \bar{q} \gamma\gamma$.
Because of its importance,
$Z \gamma \gamma$ events estimated with CompHEP have been cross-checked
at the generator level with KORALZ 4.2 \cite{Jadach} and found to
be in good agreement in the $\gamma \gamma$ invariant mass
range relevant for this study.
  
\begin{figure}[t] %
\begin{minipage}[b]{18cm}
{\epsfig{file=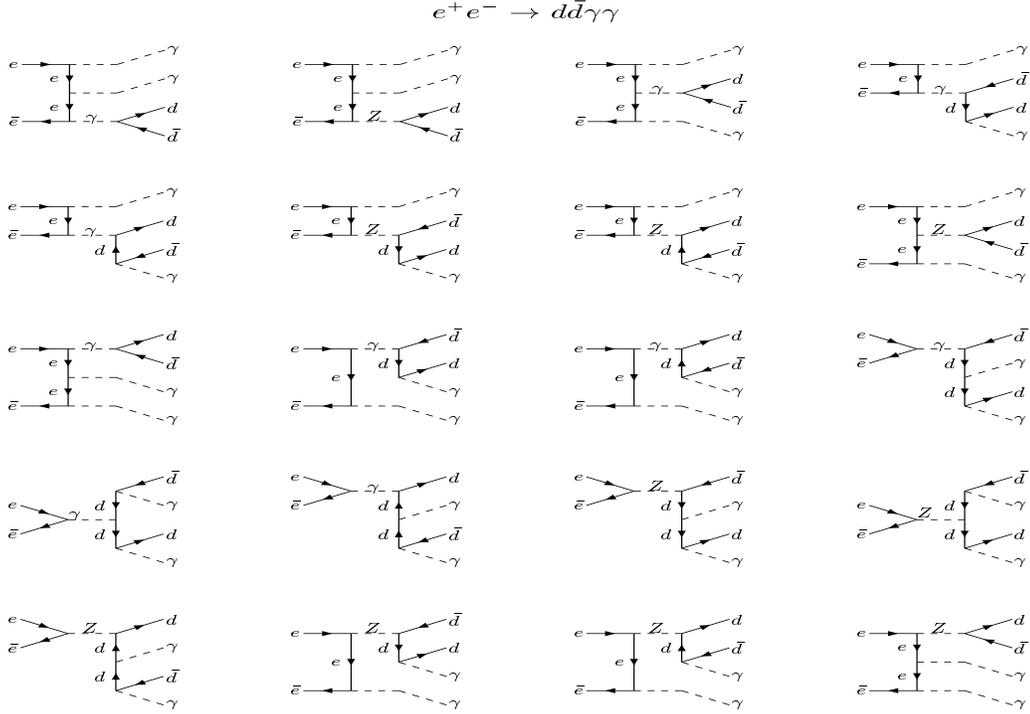,height=16cm,width=18cm}}
\vspace{-5.5cm}
\caption{Background diagrams for the reaction
$e^+e^- \rightarrow d \bar{d}\gamma \gamma$.}
\end{minipage}
\end{figure}

For the signal events in reaction (2), with contributions from
WW fusion and Higgsstrahlung processes (and small interferences
between them), we expect a signature of two photons, producing two
large electromagnetic neutral showers in the detector with no other
activities, and large missing energy due to the two undetected
neutrinos. The background diagrams contributing to reaction (2) are
shown in Fig.~2.
 They were accounted for
at the same level as the signal events. Again, the
most serious, irreducible background was found to arise from the
double bremsstrahlung process $e^+e^- \rightarrow Z \gamma \gamma
\rightarrow \nu\bar{\nu} \gamma \gamma$.

\begin{figure}[t] %
\begin{minipage}[b]{18cm} 
{\epsfig{file=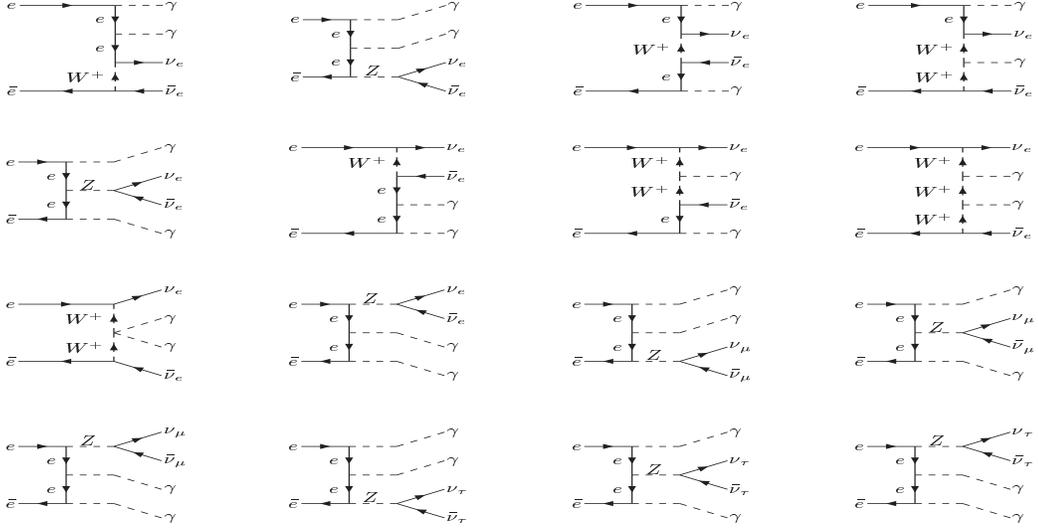,height=16cm,width=18cm}}
\vspace{-6.5cm}
\caption{Background diagrams for the reaction
$e^+e^-\rightarrow\nu\bar{\nu}\gamma \gamma$.}
\end{minipage}
\end{figure} 

Possible reducible backgrounds to \ee$\rightarrow HZ$ events which
might mimic the signal, such as the reactions
\ee$\rightarrow ZZ $ and \ee$\rightarrow WW $ with large cross sections,
 were found to be very small
after application of selection criteria.

Processes like \ee $\rightarrow \gamma \gamma (\gamma)$ or
\ee $\rightarrow$ (\ee) $\gamma \gamma$, when both electrons are
undetected, might constitute
a significant background to \ee$\rightarrow ZH$, $\nu \bar{\nu}H$
 $\rightarrow \nu \bar{\nu}\gamma
\gamma$ events. However, after kinematical
cuts their rates were also found to be small or negligible.

As the cross sections for the
background discussed are orders of magnitude larger than the
signal cross sections, we first applied the following principal cuts
at the generation level, to both the signal and
background events:

\begin{itemize}
\item the transverse energy, $E_T$, for each photon exceeds 20 GeV;
\item the two-photon invariant mass, $M_{\gamma \gamma}$, is required to be
  larger than 100 GeV;
\item the $q \bar{q}$ invariant mass, $M_{q\bar{q}}$, is
  within $M_Z \pm 20$GeV.
\end{itemize}

After these criteria, practically all Higgs events
survive, while background contributions are
substantially reduced.

 The detector response for all signal and the remaining background
 events was simulated with the parametrized detector simulation
 package SIMDET \cite{SIMDET} using parameters
 as presented in the Conceptual Design Report \cite{CDR}.

\vspace{1.0cm}
\section { BF(\hgg) measurement at 350 GeV } 
  
 Different Higgs event rates are expected
from processes (3) and (4)
 at our two energies, $\sqrt{s}$ = 350 and 500 GeV.
 The Higgsstrahlung cross
 section scales, after a maximum close its threshold, with $1/s$,
 while the fusion cross section rises logarithmically with $\sqrt{s}$.
 At 350 GeV, the Higgsstrahlung cross section
 is approximately 140 fb
 for $M_H$ = 120 GeV, which is about four times larger
 than the WW fusion cross section.
  Since the Z boson hadronic decay is by far dominant,
  $q \bar{q} \gamma \gamma$ events constitute
 the main source of the Higgs signal at 350 GeV.
  The Z invisible $\nu \bar{\nu}$ decay in the Higgsstrahlung and
  the fusion channels lead to identical event topologies
 - two isolated high energy photons
  and a large missing energy due to the two undetected neutrinos -
  are treated together in our study.

 Considering only Z
 branching fractions into $q \bar{q}$ and $\nu \bar{\nu}$ and the Higgs
 decay into two photons, we expect about
 220 respectively 130 Higgs events in reactions (1) and (2)
 at 350 GeV for an integrated luminosity of 1 ab$^{-1}$.
 Different event selection procedures were applied
to the $q \bar{q} \gamma \gamma$ and $\nu \bar{\nu} \gamma \gamma$
 event samples in order to account
for their distinct properties in the final state.

In order to avoid large not useful event samples the following
preselection procedure was applied to the 2-jet 2-photon candidates:

\begin{itemize}
\item at least two isolated neutral clusters compatible with
       photons exist, with transverse energy E$_T >$ 20 GeV for each;
       
\item no particle exists in a half cone of $10^o$
       around the isolated photon directions;

\item the number of charged tracks per event exceeds 5;

\item the visible event energy, E$_{vis}$, exceeds
    0.8 $\times \sqrt{s}$ GeV;

\item the total momentum along the beam direction
      is within $\pm$ 100 GeV;

\item all particles, excluding the two selected photons,
       were forced into two hadronic jets with an invariant
       mass compatible with the
       Z boson, 70 $ <  M_{jj} < $ 110 GeV.
\end{itemize}

 After all cuts the background to
 the Higgsstrahlung events close to $M_H$ was significantly reduced,
 but still one order of magnitude larger
 than the signal, so further and more stringent
 selection criteria were necessary to improve the
 signal-to-background ratio.

 In a first trial, we applied the conventional method
 of using consecutive cuts on kinematical variables. In particular,
 we demanded
\begin{itemize}
\item the energy of the two-photon system is between
    $ 140 < E_{\gamma \gamma} < 200$ GeV;
\item the transverse energy of the two-photon system is larger than
      50 GeV;
\item each photon polar angle is restricted to
      $|cos\theta_{\gamma}| < 0.9$;
\item the polar angle of the two-photon system is required to be
     $|cos \theta_{\gamma \gamma}| < 0.8$.
\end{itemize}

These cuts gave a selection efficiency of 56$\%$ for Higgs signal events.

 Secondly, we applied a more sophisticated selection procedure.
 Kinematical variables of the final state photons and of the \gaga
 subsystem
 were combined into a global discriminant variable~$P_H$\footnote {In
 particular, the quantities used are: energies, transverse
 energies and polar angles of both
 photons, the angle between the photons and the energy, the transverse
 energy and the polar angle of the $\gamma\gamma$ system}.
 This quantity can be considered as a measurement
 of the "Higgs-likeness" of an event, with 0$\le P_H \le$ 1.
 Background events are preferably  distributed at low $P_H$ values while
 for Higgs signal events $P_H$ is close to 1.
 The distribution of~$P_H$ is shown in Fig.~3, and a cut of
 $P_H>0.85$ was applied to select the Higgs candidates.
This method results to a signal selection efficiency of 42$\%$ and
 2.3 times less background, so that a significant better
 signal-to-background ratio exists compared to
 the application of consecutive cuts. Therefore, only results
using the discriminant variable procedure are discussed in the following.
The $\gamma \gamma$ invariant mass spectrum for
$q \bar{q} \gamma \gamma$ signal and surviving
background events is shown in Fig.~4a. The superimposed
curve is the result of a fit to the
sum of a Gaussian, used to describe the signal, and a second order
polynomial function, which was found to
describe the background in very
good approximation between 110 and 130 GeV.  
\begin{center}
\begin{figure}[t]
\begin{minipage}[b]{14cm}
\epsfig
{file=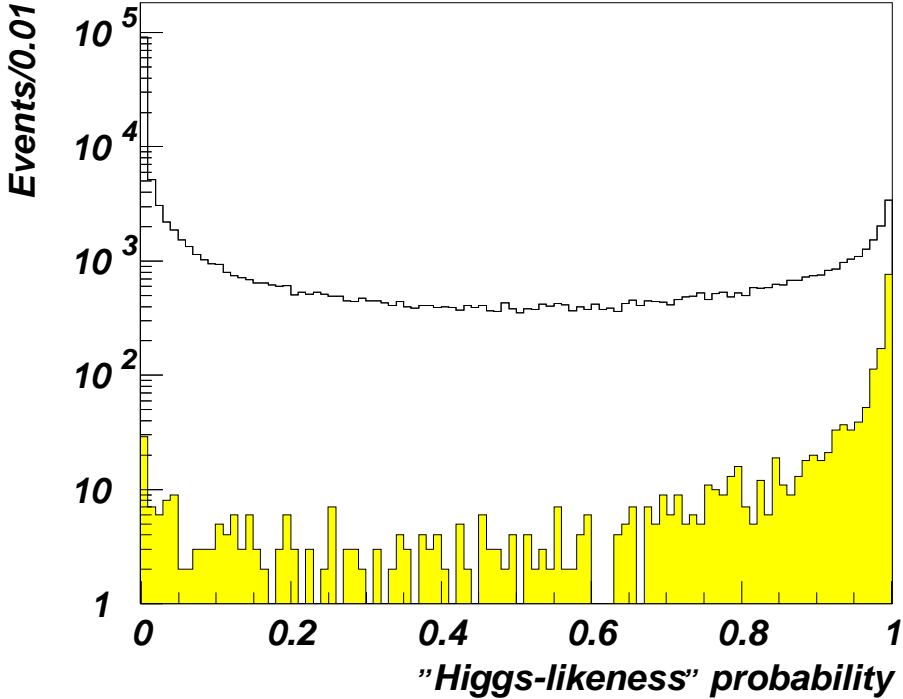,width=14cm}
\caption{"Higgs-likeness" probability for
 \ee $\rightarrow HZ \rightarrow q \bar{q} \gamma \gamma$ events (shaded)
          and the background considered.}
\end{minipage}
\end{figure}
\end{center} 
%
\begin{figure}[t] 
\begin{minipage}[t]{18cm}
\epsfig{file=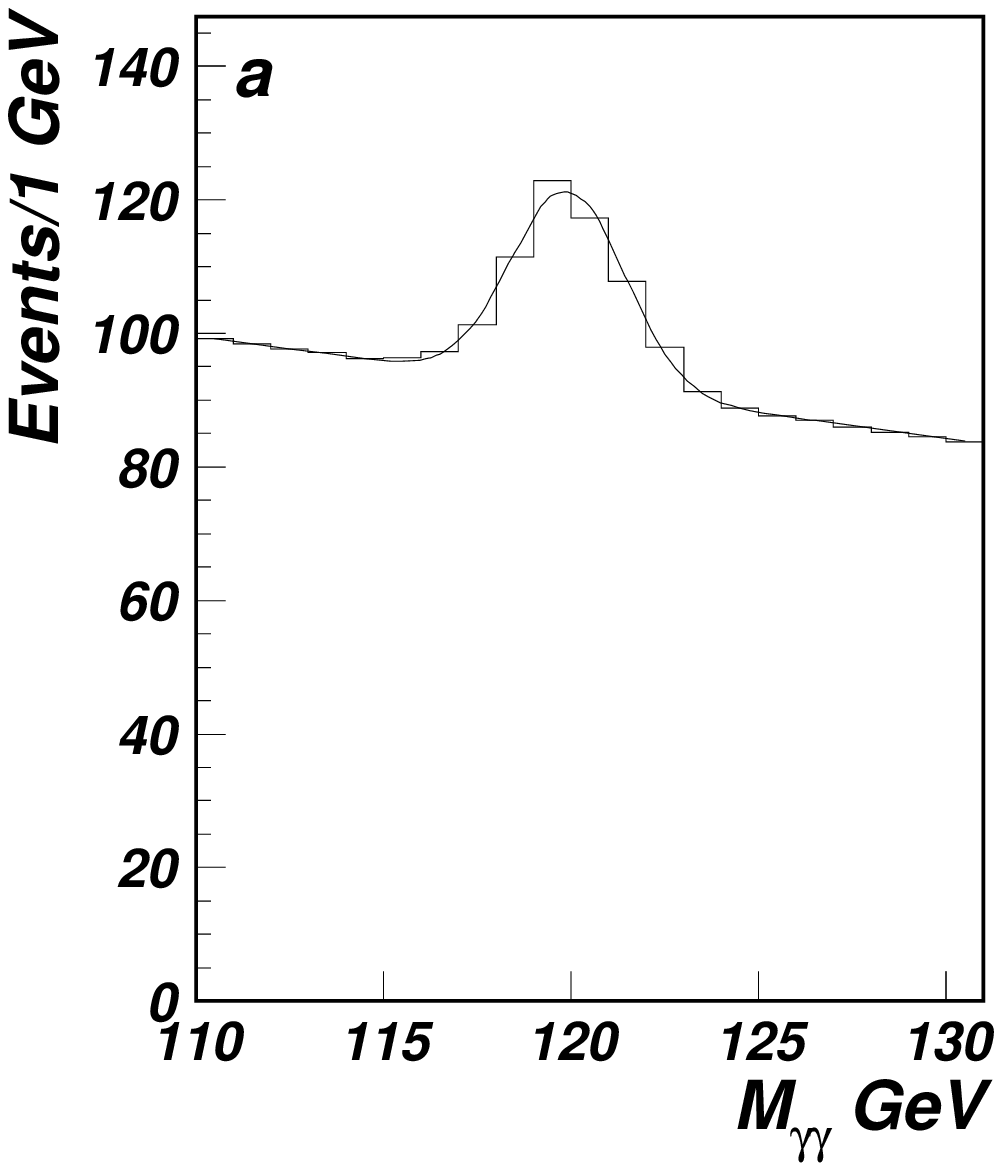,width=8cm}
\epsfig{file=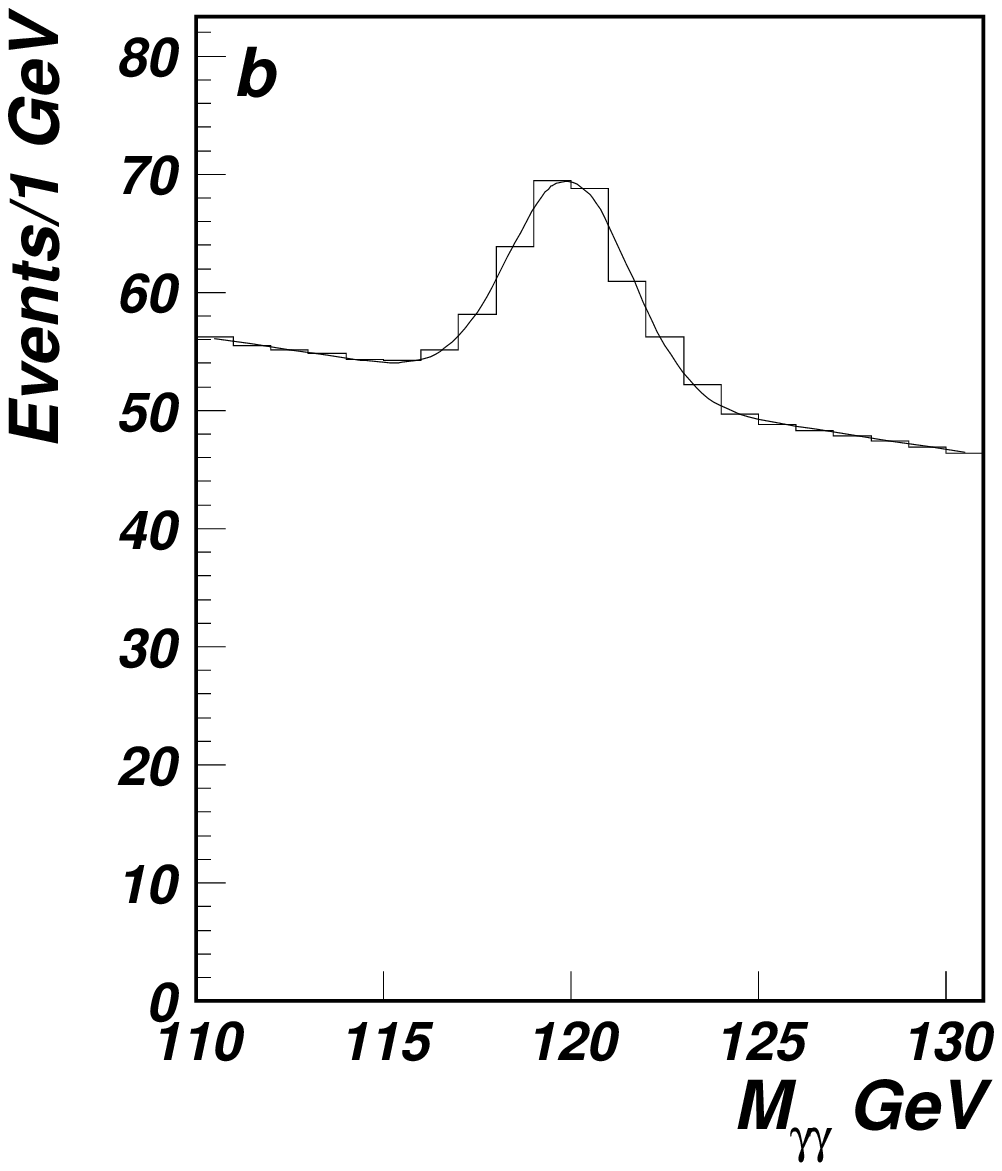,width=8cm}
\caption{ $M_{\gamma \gamma}$ invariant mass distributions for 350 GeV:
     a) $q \bar{q} \gamma \gamma$ and  
     b) $\nu \bar{\nu} \gamma \gamma$ events. The background in
        the histograms has been averaged to avoid accidental
        fluctuations.  }
\vspace{8mm}
\end{minipage}
\end{figure}

\vspace{-10mm}
From the normalizations of
the signal and background, which were allowed to vary, the number
of signal and background events in an optimal $M_{\gamma \gamma}$
window width of 2.5 GeV around $M_H$ is obtained\footnote {In order
to estimate the optimized number of signal to background events
for a narrow Gaussian resonance whose observed width is dominated
by instrumental effects, the mass window chosen should
be 1.2 $\Gamma_{exp}$, centered on the actual Higgs mass.}.
These numbers are collected
in Table~1 and suggest a statistical error of 27$\%$ for
$\sigma(HZ)\cdot BF(H \rightarrow \gamma \gamma$) in the reaction \ee
$\rightarrow ZH \rightarrow q \bar{q} \gamma\gamma$.

 The Higgs signal selection from the $\gamma \gamma \nu \bar{\nu}$
  final state relies on events without charged tracks in the
  detector. For the neutral clusters,
  compatible with photons, the following criteria were required:

\begin{itemize}
\item at least two electromagnetic clusters have
    transverse energies larger than 20 GeV;
\item their polar angles are within $|cos\Theta_{\gamma}| < 0.9$;
\item the polar angle of the two-photon system is restricted to
    $|cos\Theta_{\gamma\gamma}| < 0.8$.
\end{itemize}

 These cuts gave a Higgs selection efficiency of 45$\%$ and removed
possible backgrounds to a large extent. The resulting
 $M_{\gamma\gamma}$ mass distribution is shown in Fig.~4b.
The number of signal and background events, obtained from an analogous
fit procedure, are also shown in Table~1. They allow for
$\sqrt{S+B}/S \simeq 28.5\%$. After combining the
 $q \bar{q} \gamma\gamma$ and $\nu \bar{\nu} \gamma \gamma$ final states
and neglecting the presumably small Higgs cross section uncertainties,
the H $\rightarrow \gamma\gamma$ branching fraction error
 is estimated to 18$\%$ at $\sqrt{s}$ = 350 GeV.

\vspace{2.0cm}
\begin{center}
\begin{tabular}{||c|c|c|c||c|c||} \hline \hline

 Energy & \multicolumn{3}{c||}{350 GeV} & \multicolumn{2}{c||}{500 GeV}  \vline \vline
 \\ \hline\hline 
 Process   & $q\bar{q}\gamma\gamma$ & $\nu \bar{\nu}\gamma\gamma$  &   & 
   $q\bar{q}\gamma\gamma$ & $\nu \bar{\nu}\gamma\gamma$    \\    \hline
 Signal(S)          &   93   &  57   & 150   &   35   &   99  \\ \hline
 Bkg   (B)          &  366   & 206   & 572   &   410   &  163  \\ \hline
  $ S/\sqrt{B} $    &  4.9   & 4.0   &  6.3  &  1.7    &  10.2 \\ \hline
 Precision($\%$)    & 23.0   & 28.5  & 17.9  &  60.3   &  16.4 \\ \hline
\end{tabular}
\end{center}
 Table~1: Number of signal and background events estimated from fits to
 $M_{\gamma\gamma}$ spectra and the precisions expected
      for the cross section times the H $\rightarrow \gamma\gamma$
      branching fraction. \\

\vspace{1.0cm}
\section {\boldmath BF(\hgg) measurement at 500 GeV \unboldmath}  

  At $\sqrt{s}=500$ GeV, the
 cross sections for the Higgs signal reactions (3) and (4) are about equal,
 so that most of the Higgs events have the
  $\nu \bar{\nu} \gamma \gamma$ signature, with the dominant contribution
from the WW fusion process.

  The Higgs candidate selection procedures
 applied were similar to that at 350 GeV.
 For the $q \bar{q} \gamma\gamma$ final state,
 the "Higgs-likeness" probability $P_H$
 was also demanded to be larger than 0.85, resulting to the
 $M_{\gamma \gamma}$ distribution as shown in Fig.~5a.
 Signal and background event numbers obtained from the fit are also
 presented in Table~1.
 As can be seen, the signal is statistically insignificant,
 and we will not consider these events in our analysis.

 For the $\nu \bar{\nu} \gamma \gamma$ events we required
\begin{itemize}
\item the recoil mass against the two photon system,
 $M_{rec}=\sqrt{s+{M_{\gamma\gamma}}^2-2\sqrt{s}M_{\gamma\gamma}}$,
 is within 150 to 370 GeV and
\item E$_T^{min}$($\gamma$) $>$ 20 GeV and E$_T^{max}$($\gamma$) $>$ 50 GeV.
\end{itemize}

The first criteria reduces substantially the
 main background from double bremsstrahlung
$\gamma \gamma Z \rightarrow \gamma \gamma \nu\bar{\nu}$ events.
By this cut also the low rate
Higgsstrahlung \ee $\rightarrow ZH \rightarrow \nu \bar{\nu} \gamma \gamma$
 events were eliminated.
 Surviving background arises mainly from
 W-exchange diagrams shown in Fig.~2, with two photons radiated
from the beam particle(s).
 The transverse energy cut applied removes part of this background.

 The $M_{\gamma \gamma}$ distribution for the
 $\gamma \gamma \nu\bar{\nu}$ events selected is shown in Fig.~5b.
 The numbers obtained for the signal and background events as well
 as the precision for $\sigma\cdot BF(H \rightarrow\gamma \gamma$)
 $\simeq$ 16.4$\%$ are also given in Table~1.

\begin{figure}[t] 
\begin{minipage}[t]{18cm}
\epsfig{file=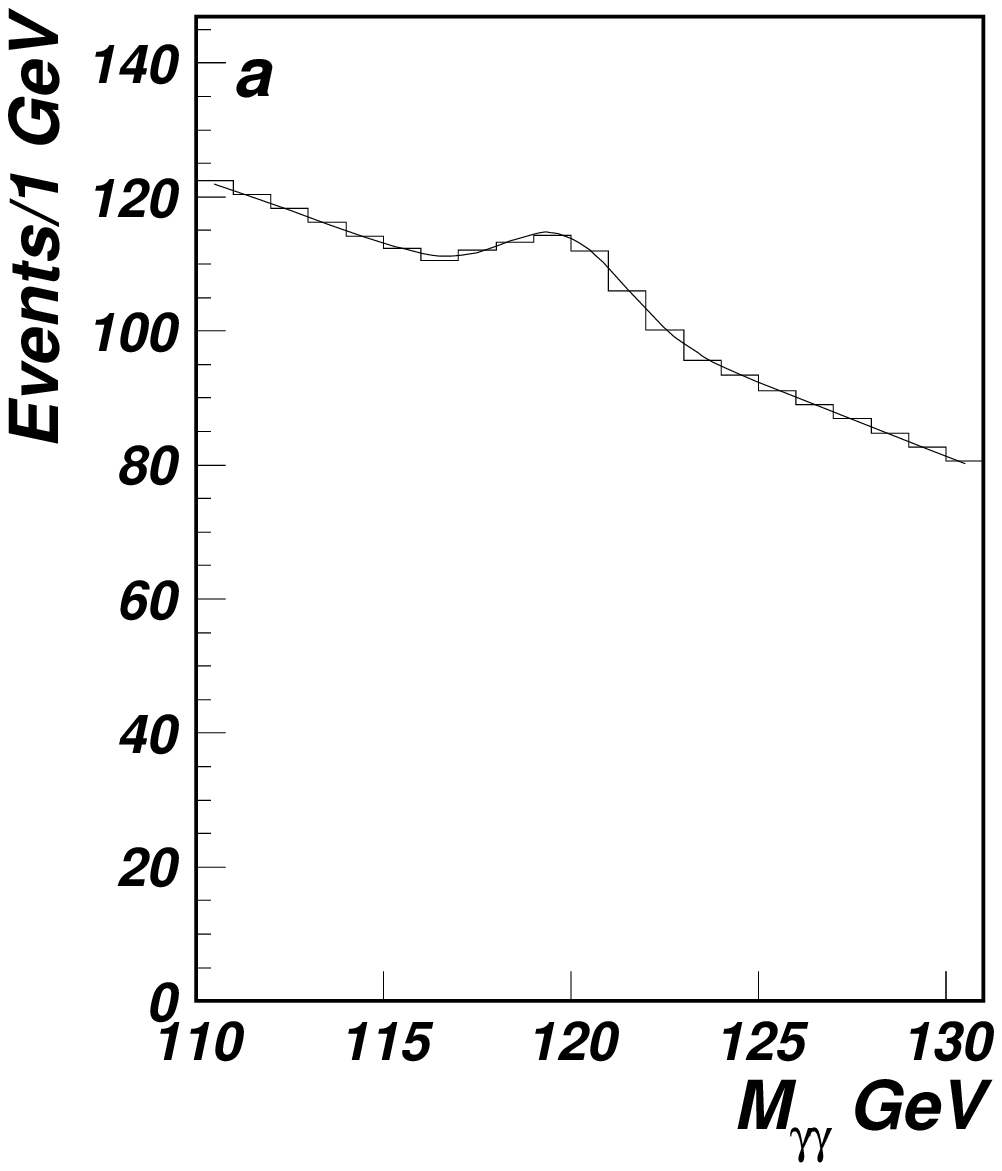,width=8cm}
\epsfig{file=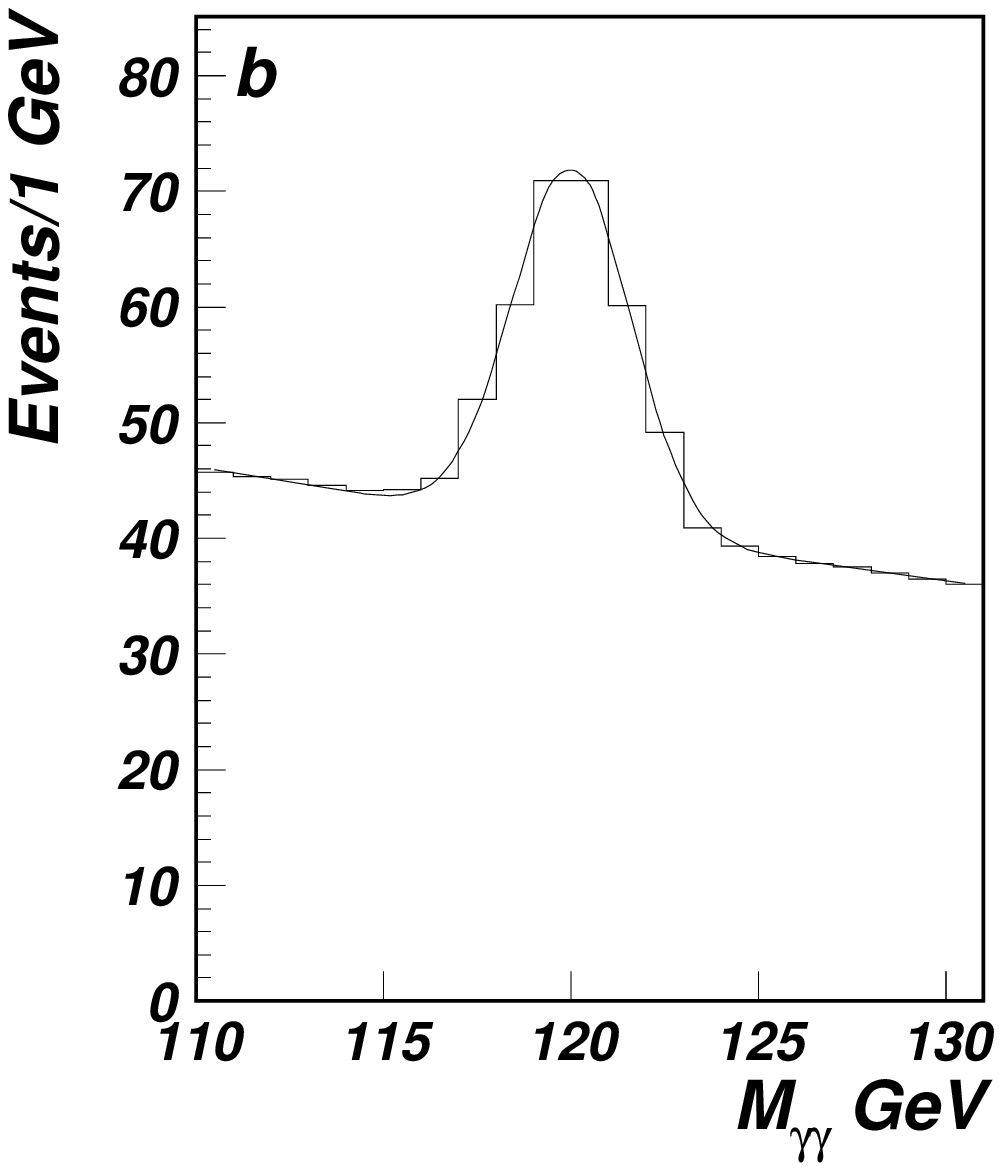,width=8cm}
\caption{ $M_{\gamma \gamma}$ invariant mass distributions for 500 GeV:
    a) $q \bar{q} \gamma \gamma$ and
    b) $\nu \bar{\nu} \gamma \gamma$ events. The background in
       the histograms has been averaged to avoid accidental 
       fluctuations. }
\end{minipage}
\end{figure}

\vspace{1.0cm}
\section {Polarization}

 Linear \ee colliders offer the possibility for longitudinal
 polarized electron and positron beams, with varying polarization degrees
 in right-handed or left-handed modes.
 Higgs boson production rates in both processes (3) and (4)
 depends significantly on the polarization degree and the
 helicity of the incoming particles.

The Higgs cross section for given electron ($P_-$) and positron ($P_+$)
polarizations normalized to the unpolarized case
can be expressed by
\begin{equation}
     R_i = 1 + \eta_i (P_-  - P_ +) -  P_-  P_ + \hspace{4mm},
\end{equation}
   where $\eta_i$ is the asymmetry factor for the cross sections,
 $\eta_i =(\sigma_i^{+-} - \sigma_i^{-+})/(\sigma_i^{+-} + \sigma_i^{-+})$,
with different indices i to indicate the Higgsstrahlung (i=1)
and the WW fusion (i=2) reactions.
  $\sigma_i^{+-}(\sigma_i^{-+})$ denotes the cross section for 100$\%$
  right-handed (left-handed) polarized electrons in collision with
 100$\%$ left-handed (right-handed) polarized positrons.
  The asymmetry factors were calculated with CompHEP
  to $\eta_1 = -0.21$ and $\eta_2 = -1$. For the WW fusion process
  the absolute value of the asymmetry $ \eta_2 $ is maximal, as this
  process occurs only via a selected combination of electron
  and positron helicities. If only the e$^-$ beam is
  polarized, $R_i = 1 + \eta_i P_-$. Some $R_i$ values
  for various left- and right-handed beam particle polarization degrees
  are shown in Table~2,
 which  illustrates the potential of polarized
 colliding beams for Higgs boson physics at a linear collider.
 Obviously, only left-handed e$^-$ which collide with right-handed e$^+$
 with largest polarization degrees as possible enhance
 the Higgs event rates at most. 
 
However, the dominant background in processes (1) and (2), such
as $ Z \gamma\gamma$ and W-exchange $\nu_e \bar{\nu_e} \gamma\gamma$, scale
in approximately the same way with beam polarizations 
as the signal processes, as was verified by CompHEP simulations.
Hence, the $\gamma \gamma$ invariant mass spectra shown
 in Figs.4 and 5 can be rescaled by the appropriate ${R_i}$ factor,
 and the statistical precision of the H $\rightarrow \gamma \gamma$
branching fraction is improved by only a factor $\sqrt{R_i}$.
  
 For the idealized case of collisions of perfect left-polarized
e$^-$ with perfect right-polarized e$^+$ beams the precision achievable
for $\sigma(H)\cdot BF(H \rightarrow \gamma \gamma)$ is expected to be
10.3$\%$  and 8.2$\%$ at 350 and 500 GeV, respectively, after combining
reactions (3) and (4) at 350 GeV and considering 
only reaction (4) at 500 GeV.
For the feasible (ambitious) case of collisions
of 80$\%$ left-handed electrons with 40 (60)$\%$ rigth-handed positrons
\cite{ecfa_desy7} the statistical precision for
$\sigma(H)\cdot BF(H \rightarrow \gamma \gamma)$ = 12.8 (12.1)$\%$,
 respectively,
10.2 (9.6)$\%$ at $\sqrt{s}$ = 350 and 500 GeV, for 1 ab$^{-1}$
integrated luminosity.
The errors of the diphoton branching fraction of the Higgs boson
are then deduced after convolution with the uncertainties
for the inclusive Higgs production rates which are expected
to be close or better~than~2\%~\cite{desch}.

\vspace{1cm}
\begin{center}   
\begin{tabular}{|c|c||c|c|} \hline
 $e^-$ beam $(P_-)$  &
 $e^+$ beam $(P_+)$  & $ e^+e^- \rightarrow HZ$  &  
$ e^+e^- \rightarrow \nu_e \bar{\nu}_e H$  \\
\hline \hline
 +1    &    0    &   0.79  &     0             \\  
 -1    &    0    &   1.21  &     2             \\  \hline
 +0.8  &    0    &   0.83  &     0.2           \\
 -0.8  &    0    &   1.17  &     1.8           \\  \hline
  +1    &    -1   &   1.58  &     0            \\
  -1    &    +1   &   2.42  &     4            \\  \hline
 +0.8  &   -0.4    &   1.07  &     0.12        \\
 -0.8  &   +0.4    &   1.57  &     2.52        \\  \hline
 +0.8  &   -0.6    &   1.19  &     0.08        \\
 -0.8  &   +0.6    &   1.77  &     2.88        \\  \hline \hline

\end{tabular}
\end{center}
Table~2: Cross section scaling factors of Higgsstrahlung
 and WW fusion processes for various left-handed and
 rigth-handed polarization degrees,
 normalized to the unpolarized event rates. \\

\vspace{1.0cm}
\section{Conclusions}

In this study we examined the prospect at a future linear
\ee collider of measuring the branching fraction 
of a Standard Model-like Higgs boson
into two photons, BF(\hgg).
A Higgs boson mass of 120 GeV and an integrated luminosity
of 1 ab$^{-1}$ at either $\sqrt{s}$ = 350 or 500 GeV were assumed.
In order to estimate the precision attainable on
BF(\hgg), all  expected backgrounds were included in the analysis,
and acceptances and resolutions of a linear collider detector were
convoluted. In particular, by simulating the 2-to-4 particle reactions
\ee $\rightarrow q \bar{q}\gamma\gamma$ and
\ee $\rightarrow \nu \bar{\nu}\gamma\gamma$, in which the
signal reactions are embedded,
the complete irreducible background has been accounted for.

 At 350 GeV, where both the Higgsstrahlung and the WW
fusion mechanisms contribute significantly, the statistical
error for BF(\hgg) is~18\%, after combining both Higgs production 
channels and and
convolution with the uncertainty of the more easily measurable inclusive
Higgs boson cross sections. It should be noted that the isolation
of \ee $\rightarrow ZH \rightarrow q \bar{q}\gamma\gamma$ signal events
required a multidimensional analysis
on a likelihood estimator. Otherwise,  background from double bremsstrahlung
is overwhelming and greatly hinders
the BF(\hgg) measurement.

At 500 GeV, only the $\nu \bar{\nu}\gamma\gamma$ final state
is worth consideration and the
application of consecutive cuts on kinematical variables resulted
in a reasonable signal-to-noise ratio and a convincing signal. 
The relative precision expected for
the $ H \rightarrow \gamma\gamma$ branching fraction
is found to be~16\%.

If 80$\%$ left-handed electrons collide with 40 (60)$\%$
right-handed positrons the Higgsstrahlung and WW fusion cross sections
are significantly enhanced, so improving substantially
the precision on BF(\hgg), even taking into account that the background
 scales in the same way.
Under such circumstances,
the uncertainty for BF(\hgg) is lowered to 12.8 (12.1)$\%$
 and 10.2 (9.6)$\%$ at $\sqrt{s}$ = 350 and 500 GeV, respectively.
With these uncertainties it should be possible
to deduce a relative precision for the diphoton Higgs
partial width of
$\frac{\Delta\Gamma(H \rightarrow \gamma \gamma)}{\Gamma(H \rightarrow \gamma \gamma)} \simeq 13.5\%(12.6\%)$ 
and 11.1$\%(10.6\%)$
 at $\sqrt{s}$ = 350 and 500 GeV, respectively,
 if an uncertainty of 4.3\% for the total
Higgs width \cite{desch} is accounted for.
These uncertainties are about a factor 
5 worse than those expected from the reaction
$\gamma\gamma \rightarrow H \rightarrow b \bar{b}$ \cite{stefan},
measurable after conversion of an \ee collider to a Compton collider.

\vspace{1.0cm}
\section*{Acknowledgments}

E.B. and R.S. would like to thank DESY Zeuthen for the kind hospitality,
support and fruitful coorperation. The work of E.B. was partly
supported by the RFBR-DFG 99-02-04011, RFBR 00-01-00704,
Universities of Russia 990588 and CERN-INTAS 99-0377 grants.



\vspace{1.0cm}
\begin{thebibliography}{99}
\bibitem{LC1} e.g. E. Accomando et al., Phys. Rep. 299 (1998) 1. 
%
\bibitem{GH}  J.F. Gonion and P.C. Martin,
 Phys. Rev. Lett. 78 (1997) 4541. 
%
\bibitem{HJS1} M. Sachwitz, H.J. Schreiber and S. Shichanin,
 DESY-123E (1997) p.449 and hep-ph/9706338; \\
M. Battaglia, Proceedings of the International Workshop
on Linear Colliders (LCWS99), Sitges, Spain, 1999. 
%
\bibitem{Bo} E. Boos, V. Ilyin, A. Pukhov, M. Sachwitz and H.J.
Schreiber, EPJ.direct C5 (2000) 1; \\
G. Borisov and F. Richard, LAL-99-26,1999, hep-ph/9905413. 
%
\bibitem{La} e.g. J.Erler and P.Langacker, hep-ph/9809352 (1998). 
%
\bibitem{CDR} Conceptual Design of a 500 GeV $e^+e^-$ Linear Collider
 with Integrated X-ray Laser Facility,
  edited by R. Brinkmann. G. Materlik, J. Rossbach and A. Wagner,
  DESY 1997-048, ECFA 1997-182. 
%
\bibitem{Reid} D. Reid, Proceedings of the International Workshop
on Linear Colliders (LCWS99), Sitges, Spain, 1999. 
%
\bibitem{CompHEP} E.E.Boos et al., INP MSU 94-36/358 and SNUTP-94-116,
 hep-ph/9503280; \\
 P. Baikov et al., Proc. of the Xth Int. Workshop on High Energy
 Physics and Quantum Field Theory, QFTHEP-95, ed. by B. Levtchenko
 and V. Savrin, Moscow, 1995, p.101; \\
 A. Pukhov, et. al., CompHEP user's manual, v.3.3, INP MSU 98-41.542 and
 hep-ph/9908288. 
%
\bibitem{bms} D. Schulte, private communication; \\ 
 T. Ohl, IKDA  96/13-rev., July 1996 and hep-ph/9607454-rev. 
%
\bibitem{HDECAY} A. Djouadi, J. Kalinowski and M. Spira,
 Comput. Phys. Commun. 108 (1998) 56. 
%
\bibitem{Jadach} S. Jadach and Z. Was, Comp. Phys. Com. 36 (1985) 191; \\
S. Jadach, B.F.L. Ward and Z. Was, Comp. Phys. Com. 66 (1991) 276. 
%
\bibitem{SIMDET} M. Pohl and H.J. Schreiber, DESY 99-030, March 1999. 
%
%
%
\bibitem{ecfa_desy7} N. Walker, contribution to the ECFA/DESY workshop,
Hamburg, September 22-25, 2000. 
%
\bibitem{desch} K. Desch, contributions to the ECFA/DESY workshop,
Obernai, 16-19 Oct. 1999 and the International Workshop
on Linear Colliders (LCWS2000), Fermi National Accelerator Laboratory,
October 24-28, 2000.
%
 \bibitem{stefan} G. Jikia and S. Soldner-Rembold, Nucl.Phys.Proc.Suppl.
82 (2000) 373; \\ Proceedings of the Workshop on Physics and
Detectors for a Linear Collider, Sitges, Spain, 29 April - 5 May 1999 and
hep-ph/9910366; \\
M. Mellis, hep-ph/0008125. 

\end{thebibliography}
\end {document}